\documentclass[letterpaper, 10 pt, conference]{ieeeconf}

\IEEEoverridecommandlockouts
\usepackage{amsmath, amsfonts}
\usepackage[caption=false]{subfig}
\usepackage{hyperref}
\usepackage{graphicx}
\usepackage{cite}
\usepackage{cleveref}
\usepackage{siunitx}
\usepackage{multirow}

\title{\LARGE \bf Human Implicit Preference-Based Policy Fine-tuning \\for Multi-Agent Reinforcement Learning in USV Swarm}

\author{Hyeonjun Kim$^\star$, Kanghoon Lee$^\star$, Junho Park, Jiachen Li, Jinkyoo Park$^\dag$
    \thanks{$^\star$Both authors contributed equally to this research.}
    \thanks{$^\dag$Corresponding author.}
    \thanks{H. Kim is with the Korea Military Academy (KMA), Seoul, Korea. {\tt\small hyunjoon0605@kma.ac.kr}. This work was done while H. Kim was with KAIST.}
    \thanks{K. Lee, J. Park, and J. Park are with the Korea Advanced Institute of Science and Technology (KAIST), Daejeon, Korea. {\tt\small \{leehoon, junho9095, jinkyoo.park\}@kaist.ac.kr}.}
    \thanks{J. Li is with the University of California, Riverside, CA, USA. {\tt\small jiachen.li@ucr.edu}.}
}

\begin{document}

\maketitle

\begin{abstract}
Multi-Agent Reinforcement Learning (MARL) has shown promise in solving complex problems involving cooperation and competition among agents, such as an Unmanned Surface Vehicle (USV) swarm used in search and rescue, surveillance, and vessel protection. However, aligning system behavior with user preferences is challenging due to the difficulty of encoding expert intuition into reward functions. To address the issue, we propose a Reinforcement Learning with Human Feedback (RLHF) approach for MARL that resolves credit-assignment challenges through an Agent-Level Feedback system categorizing feedback into intra-agent, inter-agent, and intra-team types. To overcome the challenges of direct human feedback, we employ a Large Language Model (LLM) evaluator to validate our approach using feedback scenarios such as region constraints, collision avoidance, and task allocation. Our method effectively refines USV swarm policies, addressing key challenges in multi-agent systems while maintaining fairness and performance consistency.
\end{abstract}
\section{Introduction}

Reinforcement Learning (RL) has significantly advanced in various domains, including robotics \cite{andrychowicz2020learning}, autonomous driving \cite{lu2023imitation, li2024interactive}, and drug discovery \cite{zhou2019optimization}.
In particular, Multi-Agent RL (MARL) has proven effective in addressing complex real-world scenarios that demand cooperation and competition among agents \cite{lupu2021trajectory, scheikl2021cooperative, lee2023robust, zhao2022coordination}. Despite these successes, the deployment of such systems in practical applications presents challenges extending beyond performance optimization. A key issue is incorporating the tacit knowledge of domain experts. While traditional methods for designing reward functions are effective when desired behaviors are well-defined \cite{song2021autonomous}, they often fail to encapsulate the intuition and experiential insights of experts \cite{pan2022effects, skalse2022defining}. For example, an experienced air traffic controller relies on split-second judgments informed by intricate traffic patterns, which are difficult to translate into explicit mathematical rules \cite{mcgee1998future}.

\begin{figure}[t]
    \centering
    \includegraphics[width=1.\columnwidth]{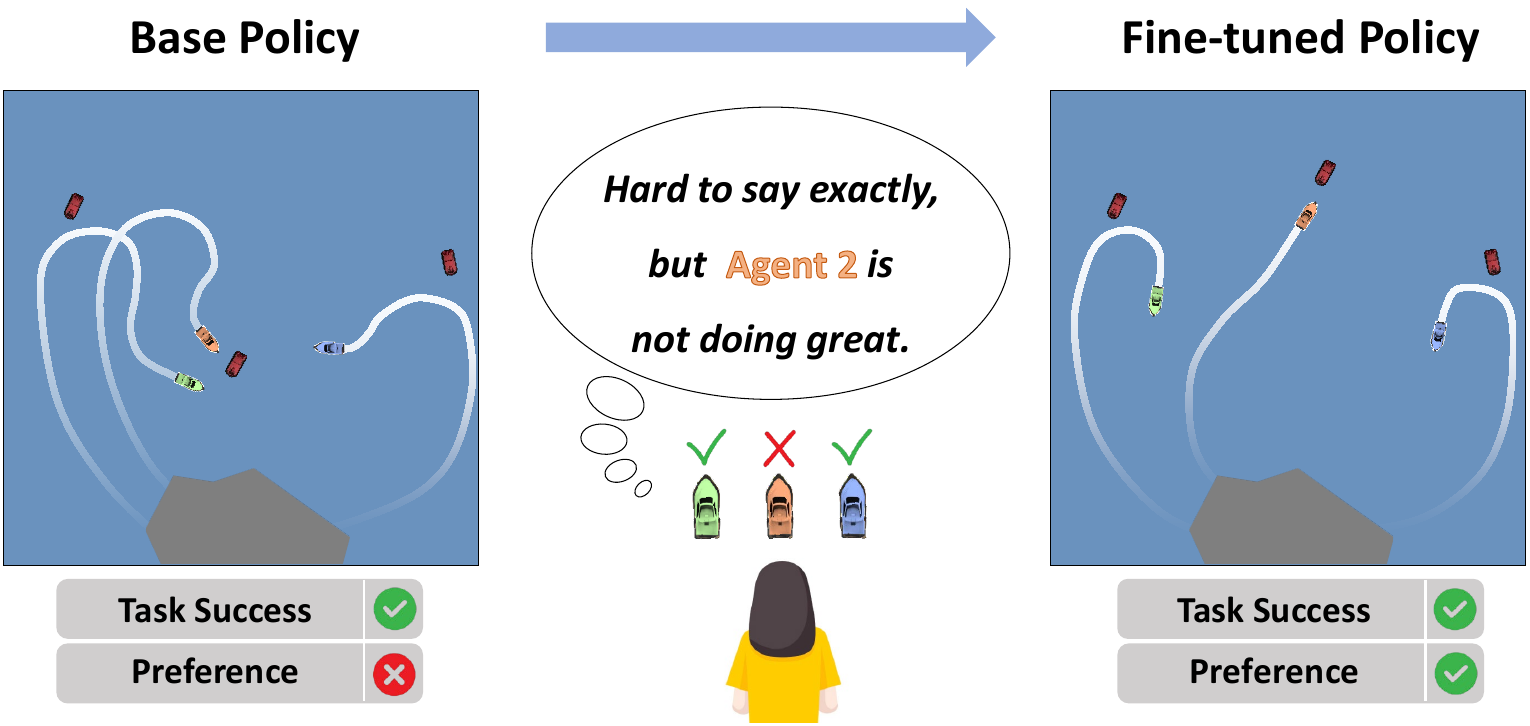}
    \caption{Illustration of the MARL policy fine-tuning process with human feedback. The base policy achieves task success but fails to satisfy preferences due to suboptimal agent performance (e.g., Agent 2). The fine-tuned policy improves both task success and preference satisfaction, demonstrating enhanced agent behavior guided by human preference.}
    \label{fig:overview}
\end{figure}

To address these challenges, RLHF has been widely adopted in robotics control \cite{christiano2017deep, ding2023learning} and large language models (LLM, \cite{stiennon2020learning, nakano2021webgpt, ouyang2022training}), leveraging human preference with reward learning. Building upon these insights, we aim to extend RLHF to refine the control of the USV swarm. USV swarm is highly versatile and can be utilized in various applications, including search and rescue missions \cite{yang2020maritime, liu2023multi}, surveillance operations \cite{shriyam2018decomposition}, and vessel protection \cite{mahacek2011dynamic}. However, their effective deployment requires overcoming challenges such as navigation, collision avoidance with static and dynamic obstacles \cite{cheng2018concise, xu2020intelligent, lin2023robust}, and achieving coordinated control \cite{almeida2010cooperative, raboin2015model}. Recently, MARL algorithms have been employed to address these challenges \cite{lee2021end, xia2023cooperative, zhang2024multiusv}. Yet, a disconnect often arises between the perspectives of model developers and end-users, making post-deployment refinement critical to align the system with user preferences. Unfortunately, most end-users lack expertise in RL, making it difficult to specify required adjustments or explicitly design suitable reward functions. To bridge this gap, we propose using RLHF to incorporate user feedback effectively and refine the model accordingly, as shown in \Cref{fig:overview}.

However, extending RLHF to multi-agent systems presents unique difficulties, as most existing MARL methods depend on team-level feedback, which hinders agent-specific guidance. For instance, \cite{zhang2024multi} investigated data coverage and proposed a reward regularization technique in MARLHF, highlighting the complexities of incorporating human feedback effectively in such systems. To address these issues, we introduced an approach that classifies agents as good or poor within a given scenario. This method resolves credit-assignment issues by establishing a direct link between feedback and agent-specific behavior. 

Our method refines a policy toward the desired direction through the following three steps: (1) Agent-Level Feedback is collected based on the trajectory information of the base policy. (2) Using the provided feedback and trajectory data, an agent-wise reward model is trained. (3) The learned reward model is integrated with the original reward to fine-tune the base policy. In our study, due to the challenges of utilizing human feedback, we developed an evaluator for diverse feedback types using LLM. Additionally, we focused on USV swarm control for the pursuit-evasion game. In summary, our key contributions are as follows:
\begin{itemize}
    \item We propose an Agent-Level Feedback system that categorizes feedback into three distinct types from a multi-agent perspective: intra-agent, inter-agent, and intra-team. This system is designed to effectively resolve the credit-assignment issue in multi-agent systems.
    \item We validate our proposed method comprehensively using an LLM evaluator with three types of feedback—region constraints, collision avoidance, and task allocation—and conduct a detailed analysis of the resulting behaviors in USV swarm control.
\end{itemize}

\section{Related Works}

\subsection{USV Swarm Control}
Unmanned Surface Vehicles (USVs) play a critical role in modern maritime operations by performing various missions reducing human risk, including search and rescue \cite{yang2020maritime, liu2023multi}, surveillance \cite{shriyam2018decomposition}, vessel protection \cite{mahacek2011dynamic}, area defense from evaders \cite{qu2023pursuit}. The achievement of these missions involves significant challenges, particularly in navigating and avoiding static-dynamic obstacles \cite{cheng2018concise, xu2020intelligent, lin2023robust}, and formation control \cite{arrichiello2006formation}. To control a swarm of USVs, sophisticated communication protocols or decentralized decision-making are essential for efficient cooperation, allowing them to tackle complex missions that require high levels of coordination and autonomy \cite{almeida2010cooperative, raboin2015model}. Incorporating deep MARL algorithms has successfully addressed these challenges \cite{lee2021end, xia2023cooperative, zhang2024multiusv}; however, MARL-based models often require modification after deployment, particularly when discrepancies arise between developer assumptions and real-world user requirements. To address this, our research proposes a method of fine-tuning the USV swarm control policy through human feedback, bridging the gap between model development and real-world application by incorporating human insights to enhance adaptability and operational effectiveness.

\subsection{Preference-based Reinforcement Learning}

Hand-designed rewards often fail to align with the true objective as they often overlook important factors and relationships within the environment \cite{hadfield2017inverse, turner2020avoiding}, which can lead to reward hacking and ultimately cause suboptimal performance \cite{pan2022effects, skalse2022defining}. To address this, Preference-based Reinforcement Learning (PbRL) grounded in human comparisons has been proposed, such as comparing two trajectories and selecting the one that better aligns with a desired objective, effectively incorporating human judgment to align rewards with human intent better \cite{christiano2017deep, ibarz2018reward}. Recent studies introduce techniques such as relabeling \cite{lee2021pebble}, bi-level optimization \cite{liu2022meta}, and meta-learning \cite{hejna2023few} in PbRL to enhance the efficiency of reward model training. LSTM has been used for non-Markovian reward modeling \cite{arjona2019rudder}, and Transformers have been utilized to model reward functions to capture trajectory-based preferences \cite{kim2023preference}. RLHF, which is PbRL with human feedback, is also extensively used to fine-tune large language models (LLMs) for tasks such as summarization \cite{stiennon2020learning}, question-answering \cite{nakano2021webgpt}, and instruction-following \cite{ouyang2022training}. Recent work has extended it to multi-agent settings, with MAPT \cite{zhu2024decoding} addressing temporal and cooperative dynamics using a cascaded Transformer. MARLHF \cite{zhang2024multi} ensures fair credit assignment and overcomes sparse feedback with reward regularization and imitation learning. Building on this foundation, our research proposes a novel preference labeling system tailored to the complexities of multi-agent systems, which can mitigate the credit assignment issue. 

\section{Problem Formulation}
In this section, we define the USV swarm pursuit-evasion game, in which USV swarms operate as both pursuers and evaders, similar to \cite{qu2023pursuit}. Evaders aim to reach a designated target without being intercepted by pursuers, who seek to protect the target by either intercepting the evader or guarding it. We formulate the game as a Partially Observable Stochastic Game (POSG, \cite{hansen2004dynamic}), which consists of $<\mathcal{I}, \mathcal{S}, \boldsymbol{\mathcal{O}}, \boldsymbol{\mathcal{A}}, \boldsymbol{\mathcal{R}}, \mathcal{T}>$ tuple, which is defined as follows:

\textit{1) Agent}: $\mathcal{I}$ is the set of agents. To facilitate description, we define the index sets of pursuers and evaders as $\mathcal{I}_\text{P}=\{1, \dots, n\}$ and $\mathcal{I}_\text{E}=\{n+1, \cdots, n+m\}$ respectively.

\textit{2) State}: $\mathcal{S}$ represents the set of states, each providing a complete description of the ongoing situation. A state includes the information for each agent $i\in\mathcal{I}$, represented by $(x^i,y^i,v^i,\theta^i,h^i)$, where $x^i$ and $y^i$ denote the $x-y$ position, $v^i$ the speed, $\theta^i$ the heading, and $h^i$ the remaining life. Also, the state incorporates the position of the target area.

\textit{3) Observation}: $\boldsymbol{\mathcal{O}}=\times_{i\in\mathcal{I}}\mathcal{O}^i$ is the joint observation space for all agents. Each agent is limited to observing other agents within its observation range, with access only to their $x-y$ positions and headings.

\begin{figure*}[t]
    \centering
    \includegraphics[width=0.9\textwidth]{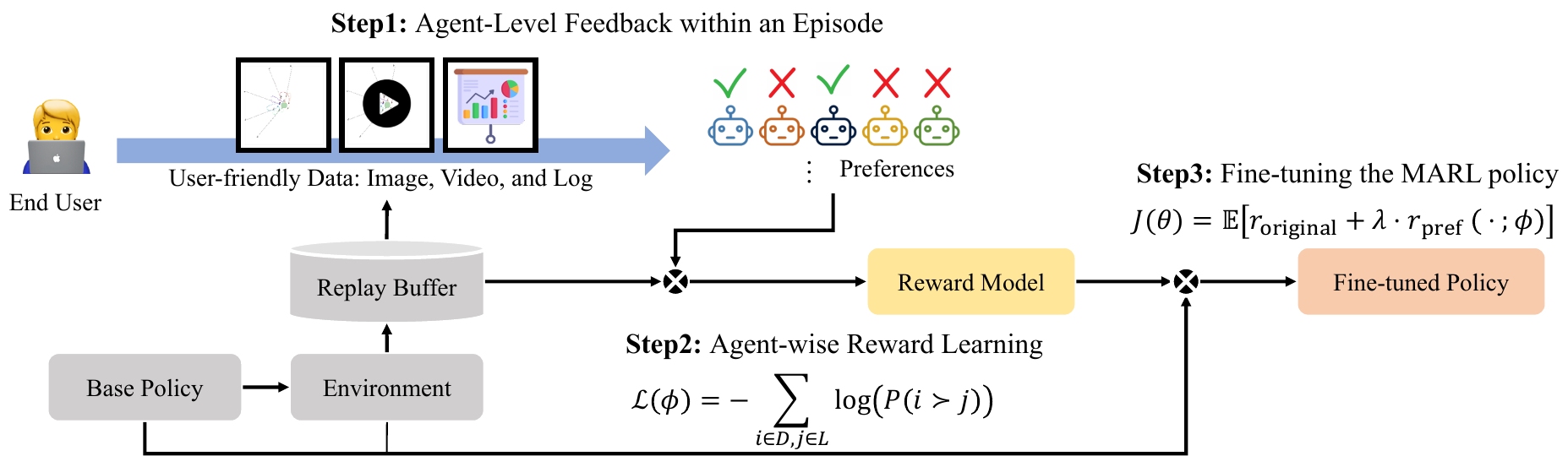}
    \caption{A diagram of the proposed method for fine-tuning MARL policies through Agent-Level human feedback.}
    \label{fig:method}
\vspace{-5pt}
\end{figure*}

\textit{4) Action}: $\boldsymbol{\mathcal{A}}=\times_{i\in\mathcal{I}}\mathcal{A}^i$ is the joint action space encompassing all agents in $\mathcal{I}$. The action of each agent corresponds to a desired relative heading, $a^i\in \mathcal{A}^i=\{-\frac{\pi}{16}, 0, +\frac{\pi}{16}\}$, which is processed by the low-level controller. It is also assumed that agents have no control over their speed.

\textit{5) Reward}: $\mathcal{R}:\mathcal{S}\times\boldsymbol{\mathcal{A}}\to\mathbb{R}^{n+m}$ is a reward function for the each agent. We only provide the reward function for the pursuers $i\in\mathcal{I}_{\text{P}}$, as the problem follows a zero-sum:
    \begin{equation}
        r^i(s,\boldsymbol{a}) = r_{\text{win}}+r_{\text{lose}}+r_{\text{collision}}^i+\sum_{j\in\mathcal{I}_\text{E}}r_{\text{intercept}}^j,
    \end{equation}
where $r_\text{win}$ denotes the reward for a successful interception of all evaders, and $r_\text{lose}$ represents the penalty incurred when any evader successfully reaches the target. $r_\text{collision}^i$ corresponds to the reward associated with a pursuer colliding with the target, while $r_\text{intercept}^j$ accounts for the reward earned when a pursuer successfully intercepts an evader $j$. Notably, the reward is shared among all pursuers except the collision reward.

\textit{6) Transition}: $\mathcal{T}:\mathcal{S}\times\boldsymbol{\mathcal{A}}\to\mathcal{S}$ is a transition function that updates the state based on the actions of all agents and the state. The position of each agent $i\in\mathcal{I}$ is updated by the following kinematic equations:
    \begin{equation}
        \begin{gathered}
            x^i_{t+1}=x^i_t+v^i_t \cdot \cos(\theta_t^i)\cdot \Delta t,\\
            y^i_{t+1}=y^i_t+v^i_t \cdot \sin(\theta_t^i)\cdot \Delta t,\\
        \end{gathered}
    \end{equation}
where $\Delta t$ denotes the time interval of the simulation. The low-level controller adjusts the heading to align with the desired heading. Agents can reduce the life of opponents within their intercept range by one unit per timestep, and any agent whose life reaches zero is terminated. Any agent that collides with obstacles is also immediately terminated.

\section{Methods}

In this section, we present our approach for incorporating human feedback into MARL to enhance policy fine-tuning. The proposed method consists of three key components as shown in \Cref{fig:method}: (1) Collecting agent-level feedback using trajectory data from the base policy, presented through user-friendly formats such as images, videos, and logs; (2) Training an agent-wise reward model based on the feedback and trajectory data, using a Bradley-Terry model; and (3) Fine-tuning the base policy by integrating the learned reward model with the original reward to produce a fine-tuned policy that aligns with human preferences while maintaining performance on the original task.
Our approach addresses challenges in identifying and assigning appropriate credit to individual agents in multi-agent environments by leveraging preference-based learning techniques. 

\subsection{Agent-Level Feedback within an Episode}

Existing MARL approaches for multi-agent systems typically rely on "Team-Level Feedback," where human preference is expressed by evaluating the overall team performance rather than individual agent contributions. While this method captures cooperative behavior, it lacks the granularity needed to provide direct feedback on specific agent actions. To address this limitation, we propose "Agent-Level Feedback," which allows human evaluators to directly label individual agents as good or poor within an episode. This fine-grained feedback improves credit assignment and enables more precise policy refinement in multi-agent systems.

Let $A$ denote the set of all agents in a given episode. During the evaluation, human feedback identifies a subset of lazy or poor agents, denoted as $L \subseteq A $, representing agents deemed to have made insufficient contributions toward the task objective. The remaining diligent or good agents, which are not identified as lazy or poor, form the set \( D = A \setminus L \).

Furthermore, We introduce a hierarchical classification of the agent-level feedback in a multi-agent system, as shown in \Cref{table:feedback_hierarchy}. While this classification does not change the learning process, it helps human evaluators categorize their observations more systemically, ensuring interpretability and consistency. Each feedback description is as follows:
\begin{itemize}
    \item \textbf{Intra-agent feedback} evaluates individual agent behavior, including adherence to movement constraints, operational boundaries, and basic protocols.
    \item \textbf{Inter-agent feedback} captures interactions between agents, such as collision avoidance, area coverage, and formation maintenance.
    \item \textbf{Inter-team feedback} focuses on higher-level strategic coordination, including evader response allocation, tactical maneuvers, and overall team positioning.
\end{itemize}

\begin{table}[t!]
\centering
\caption{Hierarchical Classification of \\Agent-Level Feedback in Multi-Agent Systems}
\vspace{-0.1cm}
\label{table:feedback_hierarchy}
\renewcommand{\arraystretch}{1.2}
\setlength{\tabcolsep}{8pt}
\begin{tabular}{c|l}
\hline
\multicolumn{1}{c|}{\textbf{Feedback Level}} & \multicolumn{1}{c}{\textbf{Evaluation Criteria}} \\
\hline\hline
\multirow{3}{*}{Intra-agent} & - Operational boundaries \\
                              & - Movement constraints \\
                              & - Basic protocols \\
\hline
\multirow{3}{*}{Inter-agent} & - Collision avoidance \\
                              & - Area coverage \\
                              & - Formation maintenance \\
\hline
\multirow{3}{*}{Inter-team}  & - Evader response allocation \\
                              & - Tactical maneuvers \\
                              & - Strategic positioning \\
\hline
\end{tabular}
\vspace{-10pt}
\end{table}

\subsection{Agent-wise Reward Learning}

In this subsection, we describe the training process of the reward model using preference-based feedback. We employ the Bradley-Terry model to capture pairwise preferences between agents, which allows us to identify underperforming agents, referred to as lazy agents. The reward function $\hat{r}_i$ for an agent $i$ is defined based on its trajectory $\tau^i=(o^i_0,a^i_0,...,o^i_T,a^i_T)$ as follows:
\begin{equation}
    \hat{r}^i = R(\tau^i, \boldsymbol{\tau}^{-i}; \phi),
\end{equation}
where $\phi$ represents the parameter of the reward function.

The reward function is modeled using a combination of Graph Neural Networks (GNN, \cite{battaglia2018relational}) and Gated Recurrent Units (GRU, \cite{chung2014empirical}) to effectively capture both spatial relationships among entities and temporal dynamics of agent behavior. We employ the GNN-GRU network instead of a Transformer due to its computational efficiency during MARL policy fine-tuning. Unlike Transformers, which require the entire trajectory to compute the terminal reward—leading to inefficiencies in a parallelized training environment—GNN-GRU supports efficient parallelization, with only the terminated environment requiring a low-cost MLP network inference to compute the terminal reward. 

To model these pairwise comparisons, we employ the Bradley-Terry model \cite{bradley1952rank}, which estimates the probability of preferring one agent over another. Specifically, for agents \( i \) and \( j \) with inferred rewards $\hat{r}^i$ and $\hat{r}^j$, the probability of preferring agent $i$ over agent $j$ is defined as:
\begin{equation}
P(i \succ j) = \frac{e^{\hat{r}_i}}{e^{\hat{r}_i} + e^{\hat{r}_j}}. 
\end{equation}
Using this formulation, we aim to assign higher rewards to diligent agents and lower rewards to lazy agents based on pairwise comparisons from the feedback data, thereby encouraging the model to effectively capture human-provided preferences. The learning objective for the reward model is to minimize the negative log-likelihood loss of these pairwise preferences as follows:

\begin{equation}\label{eq:loss1}
\mathcal{L}(\phi) = -\sum_{i \in D, j\in L} \log(P(i \succ j)).
\end{equation}

\subsection{Fine-Tuning the MARL Policy}

The fine-tuning process of the MARL policy refines agent behavior by leveraging the learned reward function from preference-based feedback while maintaining the original reward function to guide initial learning. Let $\pi_i(a_i|o_i; \theta)$ denote the policy for each agent $i\in A$, where $a_i$ represents the action of agent $i$, $o_i$ is the observation of agent $i$, and $\theta$ is the policy parameter. 

The goal of fine-tuning is to adjust the parameters $\theta$ to maximize the expected return based on a combination of the learned and original reward function. Formally, the objective function for the policy network can be expressed as:
\begin{equation}\label{eq:objective}
J(\theta) = \mathbb{E}_{\pi} \Big[ \underbrace{\sum_{t=0}^{T} \gamma^t r(\mathbf{s}_t, \mathbf{a}_t)}_{\text{Original Reward}} + \lambda \cdot \underbrace{\sum_{i=0}^{|A|} \gamma^T R(\tau^i, \boldsymbol{\tau}^{-i};\phi)}_{\text{Feedback Reward}} \Big],
\end{equation}
where $\gamma \in [0, 1]$ is the discount factor, $T$ is the episode length, $r(\mathbf{s}_t, \mathbf{a}_t)$ is the original reward at time step $t$, $R(\tau^i, \boldsymbol{\tau}^{-i};\phi)$ is a feedback reward for each agent $i$, and $\lambda$ denotes the weight for the feedback reward.
We employed the Independent Proximal Policy Optimization (IPPO, \cite{de2020independent}) algorithm to optimize the objective function in \cref{eq:objective}, using the pre-trained policy as the initial parameter. 

\section{Experiments}
\subsection{Experimental Setup}\label{sec:exp_a}

\begin{table}[t!]
\centering
\caption{IPPO Training Parameters}
\vspace{-0.1cm}
\label{tab:mappo_param}
\renewcommand{\arraystretch}{1.2}
\begin{tabular}{ll}
\hline
\textbf{Parameter} & \textbf{Value} \\
\hline\hline
Actor/Critic learning rate & $5\times10^{-4}$ / $1\times10^{-3}$ \\
Optimizer & Adam ($\epsilon=10^{-5}$) \\
Number of environments & $250$ \\
Total timesteps & $100$M \\
Batch / Mini-batch size & $250$K / $6.25$K \\
Clip / Entropy coefficient & $0.2$ / $0.01$ \\
Value function coefficient & $0.5$ \\
GAE ($\lambda$) / Discount factor ($\gamma$) & $0.98$ / $0.99$ \\
\hline
\end{tabular}
\vspace{-10pt}
\end{table}

We designed a maritime simulation environment to evaluate our proposed method. This environment simulates a 3.4 × 3.4 \si{km} area with a central island obstacle (radius: 120\si{m}). The environment implements a pursuit-evasion game scenario where five pursuers attempt to protect a designated area from five evaders. Pursuers operate at $25$ \si{knots}, while evaders move at 35 \si{knots}, reflecting typical speed differentials in maritime settings. The environment incorporates partial observability, with a detection range of 1.5 \si{km} and an attack range of 0.15 \si{km} for combat interactions. Each episode runs for a maximum of 300 steps with a time interval ($\Delta t$) of 1 second. For agent initialization, pursuers are evenly spaced 200 \si{m} from the center, while evaders are randomly spawned at distances between 1.65 to 1.9 \si{km} from the center. The initial heading of all agents is randomly set within the range of 0 to 2$\pi$. To simplify the environment, evaders naively rush toward the target island without employing any evasive maneuvers.

For the implementation of our approach, we utilize the IPPO algorithm, which enables the use of agent-wise learned rewards, with the configuration noted in \Cref{tab:mappo_param}. Each reward model is trained using data from 10,000 episodes. During policy fine-tuning, we establish a warm-up period of 2M timesteps for the critic network, ensuring stable learning as the objective function adapts to the newly introduced reward model. The fine-tuning process continues for 10M timestep, corresponding to 10\% of the base MARL model training. To evaluate the model, we systematically search for the optimal $\lambda$ values, which balance original task performance with human preference alignment. All experiments are conducted on a Linux workstation equipped with an AMD Ryzen Threadripper 3970X 32-Core Processor and an NVIDIA GeForce RTX 2080 Ti GPU.

While our research focuses on incorporating human implicit preferences into multi-agent systems, quantitatively evaluating whether these preferences are effectively reflected requires a systematic assessment methodology. To achieve this, we developed an automated evaluation script using ChatGPT-generated code, which systematically processes episode logs and assigns feedback labels based on predefined behavioral criteria. This ensures consistent and reproducible evaluation, allowing us to objectively measure the ability of the proposed method to learn and incorporate human-like preferences.
To validate our methods across diverse feedback levels in complex multi-agent environments, we utilize one key criterion per feedback level, as noted in \Cref{table:feedback_hierarchy}:
\begin{itemize}
    \item \texttt{InCircle} restricts pursuers within a designated area to maintain formation constraints.
    \item \texttt{Crossing} prevents pursuers from crossing paths in a risky manner to avoid collisions between pursuers.
    \item \texttt{Assignment} enforces distance based one-to-one matching between pursuers and evaders.
\end{itemize}

\subsection{Reward Model Evaluation}

\begin{figure}[t!]
\centering
\subfloat[Team-Level (\texttt{InCircle})]{%
  \includegraphics[width=0.45\columnwidth]{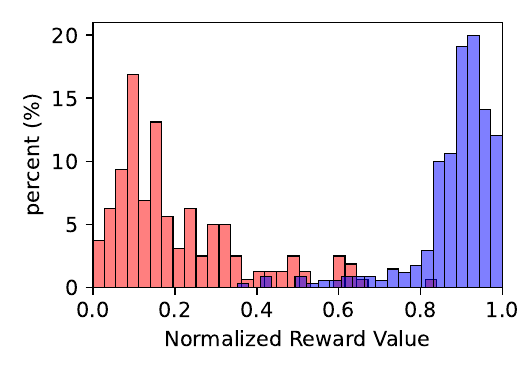}%
} \quad
\subfloat[Agent-Level (\texttt{InCircle})]{%
  \includegraphics[width=0.45\columnwidth]{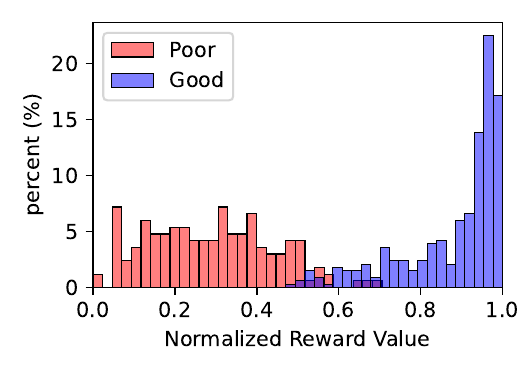}%
} \\
\subfloat[Team-Level (\texttt{Crossing})]{%
  \includegraphics[width=0.45\columnwidth]{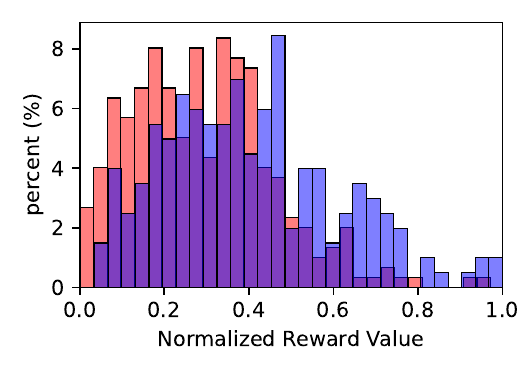}%
} \quad
\subfloat[Agent-Level (\texttt{Crossing})]{%
  \includegraphics[width=0.45\columnwidth]{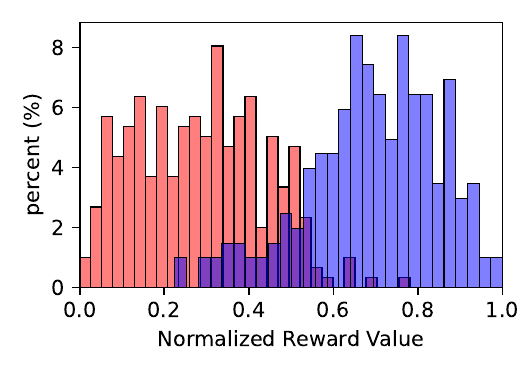}
  % {figures/distribution_agent_crossing.png}
}
\caption{Comparison of reward distribution between team-level and agent-level feedback for \texttt{InCircle} and \texttt{Crossing} scenarios.}
\label{fig:reward_dist}
\end{figure}

\begin{table}[t!]
\centering
\caption{Preference Prediction Accuracy of Reward Model ($\uparrow$)}
\vspace{-0.1cm}
\label{table:reward_model_performance}
\renewcommand{\arraystretch}{1.2}
\begin{tabular}{l|ccc}
\hline
\textbf{Methods} & \texttt{InCircle} & \texttt{Crossing} & \texttt{Assignment} \\
\hline\hline
Team-Level & $99.9$\% & $80.6$\% & $84.0$\% \\
Agent-Level & $99.9$\% & $99.3$\% & $94.7$\% \\
\hline
\end{tabular}
\vspace{-0.3cm}
\end{table}

To assess the effectiveness of agent-level feedback in reward model learning, we compare it against team-level feedback. Specifically, we aim to verify whether the agent-wise credit assignment improves preference learning and enhances reward separation between good and poor agents.

\Cref{fig:reward_dist} presents the comparison of reward distribution for both team-level and agent-level feedback in two criteria: \texttt{InCircle} and \texttt{Crossing}. In the \texttt{InCircle} scenario, team-level feedback already provides a clear distinction between good and poor agents due to the low complexity of the task. However, in the \texttt{Crossing}, where feedback involves interactions between multiple agents, team-level feedback fails to disentangle good and poor rewards effectively. In contrast, agent-level feedback produces a clearer separation, demonstrating that finer credit assignment allows for more precise differentiation between good and poor agents. As a quantitative evaluation, \Cref{table:reward_model_performance} reports the preference prediction accuracy of the reward model across the three criteria. The results show that agent-level feedback consistently outperforms team-level feedback, particularly in \texttt{Crossing} and \texttt{Assignment}, where agent-specific credit assignment is crucial. These results verify that our approach enhances reward model accuracy, leading to more reliable preference-based policy fine-tuning.

\subsection{MARL Policy Fine-tuning}

\begin{table}[t!]
\centering
\caption{Comparison of Performance and Preference}
\label{table:performance_preference_analysis}
\vspace{-0.1cm}
\renewcommand{\arraystretch}{1.2}
\setlength{\tabcolsep}{6.5pt}
\begin{tabular}{c|ccc}
\hline
\multicolumn{1}{c|}{\textbf{Criteria}} & \multicolumn{1}{c}{\textbf{Model}} & \multicolumn{1}{c}{\textbf{Performance} ($\uparrow$)} & \multicolumn{1}{c}{\textbf{Preference} ($\uparrow$)} \\
\hline\hline
\multirow{2}{*}{\centering \texttt{InCircle}} & Base & \textbf{85.30} $\pm$ 1.28\% & 67.73 $\pm$ 0.02\% \\
 & Fine-tuned & \textbf{84.73} $\pm$ 0.26\% & \textbf{72.41} $\pm$ 0.08\% \\
\hline
\multirow{2}{*}{\centering \texttt{Crossing}} & Base & 85.30 $\pm$ 1.28\% & 38.57 $\pm$ 0.02\% \\
 & Fine-tuned & \textbf{89.53} $\pm$ 0.24\% & \textbf{44.99} $\pm$ 0.30\% \\
\hline
\multirow{2}{*}{\centering \texttt{Assignment}} & Base & 85.30 $\pm$ 1.28\% & 51.63 $\pm$ 0.35\% \\
 & Fine-tuned & \textbf{88.67} $\pm$ 0.26\% & \textbf{59.86} $\pm$ 0.88\% \\
\hline
\end{tabular}
\end{table}

\begin{table}[t!]
\centering
\caption{Correlation between Original and Feedback Rewards}
\vspace{-0.1cm}
\label{table:correlation_analysis}
\renewcommand{\arraystretch}{1.2}
\begin{tabular}{c|cc}
\hline
\textbf{Criteria} & \textbf{Correlation Coefficient} & \textbf{p-value} \\
\hline\hline
\texttt{InCircle} & $-0.2195$ & $0.0282 < 0.05$ \\
\texttt{Crossing} & $+0.3139$ & $0.0014 < 0.05$ \\
\texttt{Assignment} & $+0.2048$ & $0.0409 < 0.05$ \\
\hline
\end{tabular}
\vspace{-10pt}
\end{table}

\begin{figure*}[t!]
\centering
\subfloat[Base Policy (Before fine-tuning)]{
  \includegraphics[width=0.235\textwidth]{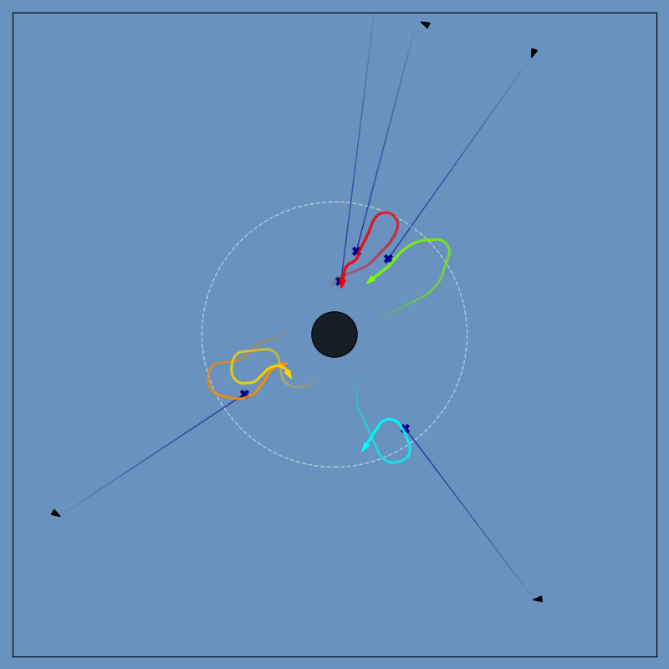}
  \label{fig:all_plots_a}
} \hspace{-7pt}
\subfloat[\texttt{InCircle}]{
  \includegraphics[width=0.235\textwidth]{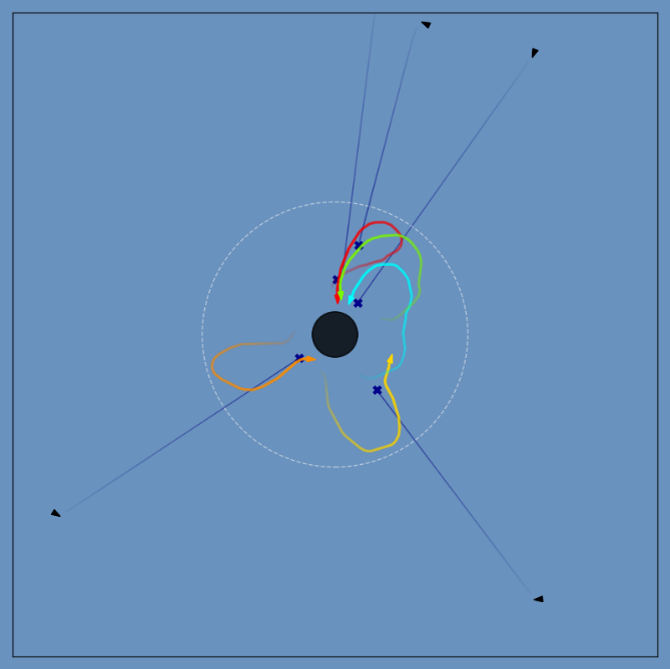}
  \label{fig:all_plots_b}
} \hspace{-7pt}
\subfloat[\texttt{Crossing}]{
  \includegraphics[width=0.235\textwidth]{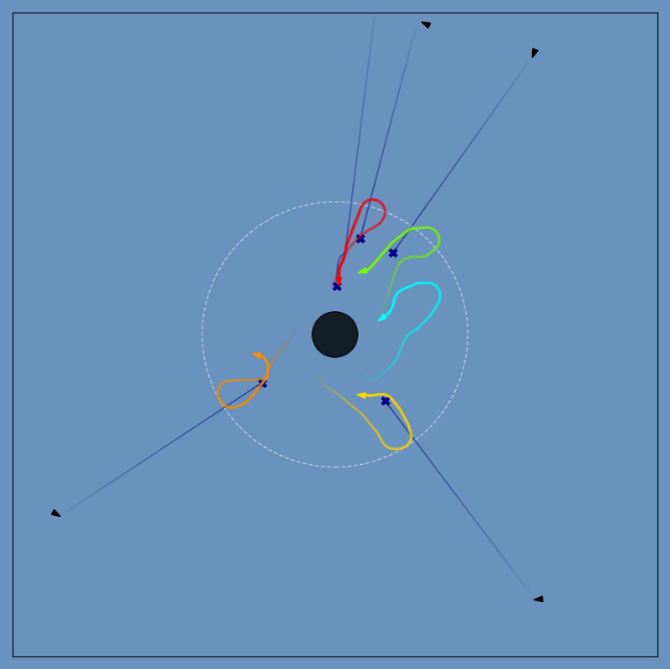}
  \label{fig:all_plots_c}
} \hspace{-7pt}
\subfloat[\texttt{Assignment}]{
  \includegraphics[width=0.235\textwidth]{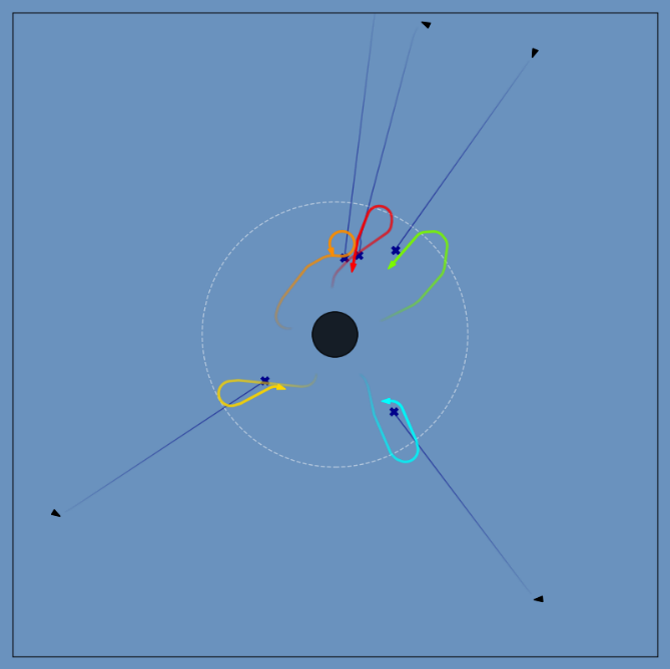}
  \label{fig:all_plots_d}
}
\caption{Comparison of USV swarm behaviors before and after fine-tuning. (a) Base policy before fine-tuning. (b)–(d) Fine-tuned policies incorporating human feedback for each criterion: \texttt{InCircle}, \texttt{Crossing}, and \texttt{Assignment}.}
\label{fig:all_plots}
\vspace{-12pt}
\end{figure*}

In this subsection, we examine the policy fine-tuning process using the learned reward model. Since team-level feedback performed poorly in reward modeling, we excluded it from this phase. To ensure consistency in evaluation, we re-utilized the evaluation script generated by ChatGPT in \Cref{sec:exp_a} to measure preference satisfaction.

\Cref{table:performance_preference_analysis} compares the task performance, which is a success rate for defending the island from the evaders, and preference satisfaction before and after fine-tuning. While the task performance remains stable or improves slightly, preference satisfaction increases consistently. It indicates that our fine-tuning approach effectively integrates human preferences without sacrificing task success. To understand why fine-tuning improves performance in certain criteria, \Cref{table:correlation_analysis} presents a correlation between the original task and learned feedback rewards.
The results show a positive correlation in \texttt{Crossing} and \texttt{Assignment}, suggesting that human preferences naturally align with task objectives in these scenarios. However, \texttt{InCircle} shows a negative correlation, indicating a potential trade-off between preference adherence and task efficiency. These findings highlight that when human preferences are well-aligned with the task objective, fine-tuning can yield performance gains that are otherwise difficult to achieve with hand-crafted rewards.

\Cref{fig:all_plots} qualitatively compares the behavior of the pursuer USV swarm before and after fine-tuning under the same initialization. In \Cref{fig:all_plots_a}, the base policy successfully defends the island, achieving task success but failing to satisfy specific criteria: (1) the green, orange, and sky-blue USVs move beyond the designated circle, (2) the orange and yellow USVs exhibit intersecting trajectories, increasing the risk of collision, and (3) the green and red USVs defend three evaders, violating the one-to-one assignment principle. After fine-tuning, \Cref{fig:all_plots_b} demonstrates that the pursuers remain within the designated circle while defending the island. \Cref{fig:all_plots_c} shows that the yellow and orange USVs maintain distinct, non-intersecting paths, mitigating collision risks. In \Cref{fig:all_plots_d}, the orange USV move upward, adjusting its trajectory to achieve a one-to-one assignment with the evaders. These results indicate that human implicit preferences can effectively drive MARL policy fine-tuning using our proposed method.

\subsection{Ablation Study}

\begin{figure}[t!]
\centering
\subfloat[\texttt{InCircle}]{
  \includegraphics[width=0.45\columnwidth]{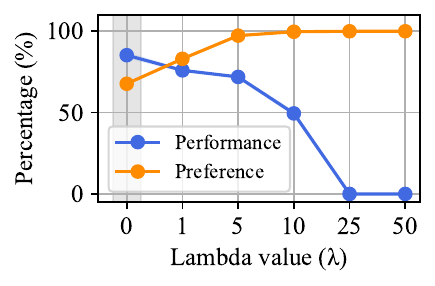}
  \label{fig:lambda_search_a}
}
\subfloat[\texttt{Assignment}]{
  \includegraphics[width=0.45\columnwidth]{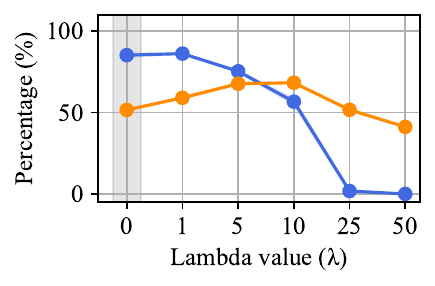}
  \label{fig:lambda_search_b}
}
\caption{Trade-off between task performance and human preference across different $\lambda$ for two criteria. The grey-shaded region indicates the base model.}
\label{fig:lambda_search}
\vspace{-12pt}
\end{figure}

In \Cref{fig:lambda_search}, we assess the trade-off between task objectives and human preferences by varying the $\lambda$ in \Cref{eq:objective}, which balances the weight for the original feedback reward. Due to space constraints, we present results for \texttt{InCircle} and \texttt{Assignment}, as \texttt{Crossing} exhibits a similar trend to \texttt{Assignment}. When the $\lambda$ is small ($\lambda \leq 5$), task performance remains relatively stable while preference increases. However, as $\lambda$ becomes large ($\lambda \geq 25$), prioritizing preference over task objectives, task performance drops to $0$\% in both criteria. Notably, in \Cref{fig:lambda_search_a}, preference increases consistently, whereas in \Cref{fig:lambda_search_b}, it initially increases but later decreases. This discrepancy arises from the correlation between each criterion and the task objective, which is negative for \texttt{InCircle} and positive for \texttt{Assignment}, as shown in \Cref{table:correlation_analysis}.

Furthermore, to assess the robustness of reward learning in terms of feedback inconsistency, we conducted experiments with varying levels of feedback noise: $0$\%, $1$\%, $5$\%, and $10$\%.
Noisy data was generated by randomly altering labels in LLM-labeled data to incorrect ones according to the specified ratio. \Cref{table:preference_accuracy_noise} shows the preference prediction of reward model for each noise level. Results indicate that our method maintains stable performance up to $5$\% noise levels, with moderate degradation observed at $10$\% noise. This demonstrates the robustness of our approach to potential variations in human feedback quality.

\begin{table}[t!]
\centering
\caption{Reward Model Prediction Accuracy for Noise Levels ($\uparrow$)}
\label{table:preference_accuracy_noise}
\vspace{-0.15cm}
\renewcommand{\arraystretch}{1.2}
\begin{tabular}{l|ccc}
\hline
\textbf{Noise Level} & \texttt{InCircle} & \texttt{Crossing} & \texttt{Assignment} \\
\hline\hline
$0$\% (Original) & $99.9$\% & $99.3$\% & $94.7$\%\\
$1$\% & $99.6$\% & $97.6$\% & $92.5$\% \\
$5$\% & $99.8$\% & $97.2$\% & $91.4$\% \\
$10$\% & $99.0$\% & $95.0$\% & $91.0$\% \\
\hline
\end{tabular}
\vspace{-10pt}
\end{table}
\section{Conclusion}

In this work, we propose a novel method for integrating human implicit feedback into MARL policy fine-tuning through agent-level feedback. This approach simplifies the feedback process for human annotators while improving reward learning and credit assignment. Additionally, we introduce a structured classification of feedback types to improve its applicability in multi-agent systems. Experimental validation across three criteria demonstrates that our method effectively aligns MARL policies with human preferences while maintaining task performance. However, our approach has a limitation in handling the inconsistency of large-scale human feedback, which can affect reward model reliability. Future work should explore strategies to reduce the dependency on extensive feedback while improving robustness, such as incorporating active learning techniques for selective feedback acquisition.

{
\bibliographystyle{IEEEtran}
\bibliography{references}
}

\end{document}